\documentclass[12pt]{article}
\usepackage{graphicx}
\begin{document}

version c5 25.9.2007
\begin{center}
{\Large Phenomenological study of two-gluon densities}
\vspace{0.3cm}

H.P. Morsch\\
Institut f\"ur Kernphysik, Forschungszentrum J\"ulich, D-52425
J\"ulich, Germany\\ and Soltan Institute for Nuclear Studies, Pl-00681
Warsaw, Poland \\P. Zupranski \\
Soltan Institute for Nuclear Studies, Pl-00681 Warsaw, Poland
\end{center}

\begin{abstract}
The decay of colour neutral 2-gluon systems in $q\bar q$ and $2q2\bar
q$ has been simulated with the Monte-Carlo method, taking into account
an effective 1-gluon exchange interaction between the emitted quarks, which was
folded with a 2-gluon density determined self-consistently. Finite 2-gluon
densities were obtained with a mean square radius $<r^2>$
of about 0.5 fm$^2$. These give a significant contribution to the 2-gluon field
correlators and the gluon propagator from lattice QCD.  \\  
By solving a relativistic Schr\"odinger equation the binding potential
of the 2-gluon system is computed. By the relation $2g\rightarrow (q\bar
q)^n$ this corresponds to the confinement potential between
quarks, which is in excellent agreement with the confinement potential from lattice QCD. 
The lowest bound state in this potential allows an interpretation of the 2-gluon
mass as binding energy of the two gluons. The lowest glueball with 
$E_o=0.68\pm0.10\ GeV$ is consistent with the scalar $\sigma$(550), with a mass 
significantly lower than obtained in lattice QCD simulations. However,
the first and second radial excitations agree with the lattice
results.\\  

PACS/ keywords: 12.38.Aw, 12.38.Lg, 12.39.Mk, 02.70.Uu/ gluon-gluon
coupling in QCD, effective 1-gluon exchange interaction, lattice gluon field
correlators and gluon propagator, confinement potential, masses of glueballs.
\end{abstract}

Quantum chromodynamics (QCD) is not well understood in the infrared
region, where the most important properties of the strong interaction
are generated, i.e.~the hadronic masses and the confinement of quarks and
gluons to hadrons. As perturbative methods are generally not applicable,
simulations of the QCD equations on the lattice~\cite{latt} together
with studies of Dyson-Schwinger equations~\cite{DSE1,DSE2} offer the best
known methods to study the structure of QCD in the infrared region. 
However, because of their complexity and severe problems in the
lattice simulations (due to discretisation, with small quark
masses and the need of a very high computing power) it is very difficult
to reveal the underlying structure of QCD. Therefore, alternative
approaches are welcome, which can shed further light on the 
underlying physics.

In the present paper we discuss a phenomenological description of two-gluon
systems following the old idea (see e.g.~in ref.~\cite{Corn}), that the
non-Abelian structure of QCD may allow a coupling of two gluons to bound
colour singlets with spin J=0 (and 2). For the description of such a scalar 
bound state $\Phi$ we may write the radial wave functions in the form 
\begin{equation}
\label{eq:phiwf}
\psi_{\Phi}(\vec r=\vec r_1-\vec r_2)=[\psi_1(\vec r_1)\ \psi_2(\vec
r_2)]\ ,
\end{equation}
where $\psi_j(\vec r_j)$ are the radial wave functions of the two gluons.
To investigate the properties of such a 2-gluon system
we studied the decay $2g\rightarrow q\bar q + 2q2\bar q$ with an
attractive interaction between the emitted quarks. 

There is evidence for the existence of scalar and tensor gluonic
fields from the formation of glueballs in pure Yang-Mills theory~\cite{glue},
from the mechanism of Pomeron-exchange~\cite{pom,pom1} in high energy
hadron-hadron scattering (which is understood by the exchange of two
non-perturbative gluons which couple mainly to spin zero), scalar
excitation of baryon resonances~\cite{MoZuneu} and hadron 
compressibilities~\cite{Mo}. Finally, the effective confining
quark-quark potential~\cite{qq} contains a sizeable scalar contribution.

Assuming an effective 1-gluon exchange interaction $V_{1g}(R)=-\alpha_s/R$ between the
emitted quarks (or antiquarks) with relative distance $R=|\vec r_i-\vec r_j|$,
the decay from a 2-gluon system $2g\rightarrow (q\bar q)^n$ requires a
modification of the free q-q interaction by the density of the 2-gluon
system, which may be expressed by a folding integral 
\begin{equation}
\label{eq:qq}
V_{qq}(R)=\int d\vec r\ \rho_{\Phi}(\vec r\ )\ V_{1g}(\vec R-\vec r\ )\ ,
\end{equation}
where $\rho_{\Phi}(\vec r\ )$ is the 2-gluon density $\rho_{\Phi}(\vec r\
)=|\psi_{\Phi}(\vec r\ )|^2$. 

It is interesting to note, that for a spherical density the Fourier transform
of eq.~(\ref{eq:qq}) to momentum (Q) space yields 
\begin{equation}
\label{eq:qqQ}
V_{qq}(Q)=-\frac{4\pi \alpha_s}{Q^2}\ \rho_{\Phi}(Q)\ ,
\end{equation}
where $\rho_{\Phi}(Q)=4 \pi \int r^2 dr\ j_o(Qr)\ \rho_{\Phi}(r)$.
Comparing this with the standard 1-gluon exchange force yields a Q-dependent
strong coupling  
\begin{equation}
\label{eq:qq1}
\alpha_s(Q)= \alpha_s\ \rho_{\Phi}(Q)\ ,
\end{equation}
which is qualitatively consistent with 
the known fact of a ``running'' of $\alpha_s(Q)$ and the condition 
$\alpha_s(Q)$$\rightarrow$ 0 for $Q$$\rightarrow$$ \infty$ (asymptotic
freedom)~\footnote{However, is is clear that the 2-gluon densities
  deduced here fall off much faster than the slow logarithmic decrease found
  for $\alpha_s(Q)$~(see~\cite{PDG}).}. Note, that the 2-gluon densities
discussed below yield  $\alpha_s(Q)$ in reasonable agreement with the lattice
data of ref.~\cite{Fu} for small momentum transfers.

Further, the interaction has to be modified by the finite size of the
decaying 2-gluon system. This is approximated in our calculation by
replacing the 1-gluon exchange potential $V_{1g}(R)$ by 
\begin{equation}
\label{eq:m1g}
V'_{1g}(R)=V_{1g}(R)\ e^{(-aR^2)}\  
\end{equation}
with the requirement $<r^2_{V'}>\approx<r^2_{\rho_{\Phi}}>$.
This effect is shown in the upper part of fig.~1, which gives a 2-gluon
density by the upper dot-dashed line and pure 1-gluon and modified 1-gluon exchange
potential~by the lower dashed and dot-dashed lines, respectively. The resulting
interaction is given by 
\begin{equation}
\label{eq:qq2}
V'_{qq}(R)=\int d\vec r\ \rho_{\Phi}(\vec r\ )\ V'_{1g}(\vec R-\vec r\ )\ ,
\end{equation}
which has the characteristcs of a quite rapid fall-off to zero for
large radii (solid line in the upper part of fig.~1).

Finally we have to take into account the fact, that the 
decay into $2q 2\bar q$ favours relative angular momentum L=0
between the emitted quarks, whereas for the decay in $q \bar
q$ the outgoing quarks are in a relative p-state (L=1). For the 
p-wave density $\rho^p_{\Phi}(\vec r\ )=\rho^p_{\Phi}(r)\ Y_{1,m}
(\theta,\phi)$ we take $\rho^p_{\Phi}(r)=(1+\beta\cdot
d/dr)\rho_{\Phi}(r)$ with the constraint $<r>\ =\int d\tau\ r 
\rho_{vec}(r)=0$ (to suppress spurious motion), by which $\beta$ is
fixed. 
 
By relativistic Fourier transformation~\cite{Kelly} the effective
interaction~(\ref{eq:qq2}) can be transformed to momentum space 
\begin{equation}
\label{eq:trans}
V'_{qq}(Q')= 4\pi \int R^2 dR\ j_o(Q R)\ V'_{qq}(R)\ ,
\end{equation}
with $Q'=Q\ \sqrt{1+[Q^2/4m_{\Phi}^2]}$ and
$m_{\Phi}$ being the mass of the 2-gluon field. For $\rho_{\Phi}(r)$ and
$\rho^p_{\Phi}(r)$ the potential in momentum space $V'_{qq}(Q)$ is given in
the lower part of fig.~1 by the solid 
and dot-dashed line, respectively, which show significant
differences. Due to the finite size effects both interactions are
finite at vanishing momentum transfer.

Using this effective q-q interaction Monte-Carlo simulations of gluon-gluon
scattering have been performed in fully relativistic kinematics using the CERN
routine GENBOD~\cite{GEN}, in which the 2 gluons in the final state
can decay in $q \bar q$ and $2q 2\bar q$ (we used massless quarks).
Random Q-transfers between 0 and 6 GeV/c have been used, from these the
momenta $\vec p_i$ of the outgoing quarks or antiquarks were also
determined randomly, but restricted by the available phase space. 
The potential  $V'_{qq}(\Delta p)$~(\ref{eq:trans})  
has been used as a weight function between the outgoing quarks (with the
relative momenta $\Delta \vec p=\vec p_i - \vec p_j$) in addition to
the normal phase space weight. Resulting gluon momentum distributions 
$d_{q\bar q}(Q)$ and $d_{2q2\bar q}(Q)$ for decay into 
$q \bar q$ and $2q 2\bar q$ were generated. Their sum
$D_{\Phi}(Q')=d_{q\bar q}(Q') +d_{2q2\bar q}(Q')$ can be
related to the radial density $\rho_{\Phi}(r)$ of the 2-gluon system  
\begin{equation}
\label{eq:distr}
D_{\Phi}(Q')=4 \pi \int r^2 dr\ j_o(Qr)\ \rho_{\Phi}(r)\ ,
\end{equation}
with $Q'$ as in eq.~(\ref{eq:trans}). The mass $m_{\Phi}$ to determine
$Q'$ in the relativistic 
Fourier transformations~(\ref{eq:trans}) and (\ref{eq:distr}) has been used as
a fit parameter, for which a value of 0.68 GeV yields the best results. 

The condition, that $\rho_{\Phi}(r)$ in the interaction~(\ref{eq:qq2})
and in eq.~(\ref{eq:distr}) should be the same, allows us to determine
this density. 
As the 2-gluon density was not known initially, our simulation
procedure had to be iterated: starting with a certain density
$\rho_{\Phi}(r)$ the folding potential~(\ref{eq:qq2})
was calculated. The Fourier transform~(\ref{eq:trans}) was then
inserted in the Monte-Carlo simulations, from which gluon 
momentum distributions $d_{q\bar q}(Q)$ and $d_{2q2\bar q}(Q)$ were
generated as shown in the upper part of fig.~2. By inversion of
eq.~(\ref{eq:distr}) a Fourier retransformation of $D_{\Phi}(Q)$ to
r-space had to be made and the resulting density $\rho_{\Phi}(r)$ 
compared to the one used in eq.~(\ref{eq:qq2}). Self-consistency of
both densities was required. 

Resulting momentum distributions $d_{q\bar q}(Q)$ and $d_{2q2\bar
  q}(Q)$ for a self-consistent solution (discussed below) are given
in the upper part of fig.~2.
These are very different for the two decays: the decay in $q \bar q$ 
falls off rapidly with increasing Q, whereas a much slower fall-off for
$2q 2\bar q$ decay is observed. This difference can be readily explained by
the fact, that 4 quarks can absorb much higher momenta than 2 quarks.
We see that the sum $D_{\Phi}(Q)$ is in reasonable agreement with
$\rho_{\Phi}(Q)$ from the Fourier transformation of $\rho_{\Phi}(r)$ inserted
in eq.~(\ref{eq:qq2}) (dot-dashed line), which is
 quite well approximated by a radial dependence  
\begin{equation}
\label{eq:wf}
\psi_{\Phi}(r)=\psi_o\ exp[-(r/a)^{\kappa}]
\end{equation} 
with values of $a$ and $\kappa$ in table~1 and $\psi_o$ taken
 from the normalisation  $4\pi\int r^2dr \ \psi_{\Phi}(r)^2$=1. 

The self-consistency condition for our simulations is far from
trivial. The resulting density is given in the lower part of
fig.~2, which indicates clearly that a self-stabilized 2-gluon
field is generated. In table~1 we give two solutions of about
equal quality, which may reflect the minor ambiguities of our method.
The dashed area indicates the estimated uncertainties of the deduced 
density. 

The extracted 2-gluon density should give a significant contribution to the
gluon 2-point functions extracted from lattice QCD simulations in form of
gluon field correlators~\cite{FC,DiG} and the QCD gluon propagator
(see e.g.~\cite{Mand}). This is shown in fig.~3. In the upper part we make a
comparison of our results with the 2-gluon field correlator $C_{\perp}(r)$  
of Di~Giacomo et al.~\cite{DiG}. $C_{\perp}(r)$ has been weighted exponentially with
distance (yielding $<r^2_{C}>\sim$0.01 fm$^2$); therefore we have to compare
the logarithmic values of $C_{\perp}(r)$ with our 2-gluon density
(lower solid line). This leads to a reasonable agreement at large radii. 
For a comparison with the gluon propagator we have taken the
lattice data (obtained in Landau gauge) of Bowman et al.~\cite{Bow04}. These
are compared with the Fourier transformed 2-gluon density $\rho_\Phi$(Q) shown
by the lower solid line in the lower part of fig.~3. 

In spite of a reasonable agreement with our densities at large radii or
small $Q$-values, respectively (note that in fig.~3 the gluon
propagator is multiplied by $Q^2$), both lattice data show deviations, which
may be described by a 2-gluon vector component (where the two gluons
couple to angular momentum L=1). Such a (current) field is constrained by the
condition $<r>=0$ to suppress spurious motion. We assume a
form $\rho_{vec}(r)=\rho_o (1-0.25\ r/a')\ exp\{-(r/a')\}$ with $a'$=0.11 fm,
which fullfills the requirement $<r>=0$, yielding a contribution given by
the dot-dashed lines in fig.~3. Using for this a strength of about 18 \% of the
scalar component yields a consistent description of the (gauge
independent) 2-gluon correlator and the gluon propagator in Landau gauge. The
latter is not very different in Laplacian gauge~\cite{gpropl}, but a fit to
the gluon propagator in Coulomb gauge~\cite{Coul} yields a relative vector
component about a factor 2 smaller, but in this gauge the gluon propagator is
not covariant. 

\begin{table}
\caption{Deduced density parameters and energies of the lowest radial
  eigenmodes with the units fm, fm$^2$ and GeV for $a$, $<$r$^2>$,
  $\delta m$ and $E_i$, respectively. } 
\begin{center}
\begin{tabular}{clcc|c|ccc}
2-gluon field& ~~$\kappa$ &$a$& $<$r$^2>$ & $\delta m$&$E_o$& $E_{1}$ &$E_{2}$\\
\hline
(solution 1)&1.50&0.66 &0.48&0.04&0.68$\pm$0.10&~1.70$\pm$0.15&2.58$\pm$0.20\\ 
(solution 2)&1.51&0.69 &0.52&0.12&0.68$\pm$0.10&~1.74$\pm$0.15&2.68$\pm$0.20\\ 
\end{tabular}
\end{center}
\end{table}

In various investigations of the gluon propagator and Pomeron-exchange
it has been concluded, that 2-gluon fields should be
massive. Cornwall~\cite{Corn} has discussed the gluon
propagator in relation to a (relativistic) generation of mass. In 
studies of Pomeron-exchange~\cite{pom,pom1} an effective gluon mass between
0.3 and 0.8 GeV has been extracted. In our relativistic Fourier
transformations we had to assume $m_{\Phi}\sim$ 0.68 GeV. This is
consistent with the gluon pole mass of $0.64\pm 0.14$ GeV deduced in
ref.~\cite{Alex}. We shall see, that the extracted mass can be understood as 
binding energy of the 2-gluon system including relativistic mass corrections.

A finite 2-gluon density as shown in fig.~2 may be interpreted as a 
quasi bound state of the two gluons (glueball). Therefore, from the 
2-gluon density the binding potential of the 2-gluon system can
be obtained by solving a three-dimensional reduction of the 
Bethe-Salpeter equation in form of a relativistic Schr\"odinger equation 
\begin{equation}
-\bigg( \frac{\hbar^2}{2\mu_{\Phi}}\ \Big [\frac{d^2}{dr^2} +
  \frac{2}{r}\frac{d}{dr}\Big ] - V_{\Phi}(r)\bigg) \psi_{\Phi}(r) =
  E_i \psi_{\Phi}(r)\ ,  
\label{eq:4}
\end{equation}
where $\psi_{\Phi}(r)$ is the 2-gluon wavefunction~(\ref{eq:phiwf}) and
$\mu_{\Phi}$ a relativistic mass parameter, which is related to the
mass $m_{\Phi}$ of the 2-gluon system by $\mu_{\Phi}=\frac{1}{4}\
m_{\Phi}+\delta m$. In the 
non-relativistic limit $\delta m=0$ and $\mu_{\Phi}$ corresponds to the
reduced mass $\mu=\frac{m1\cdot m_2}{m_1+m_2}$ which gives
$\mu=\frac{1}{4}\ {m_{\Phi}}$ for $m_1=m_2=m_{\Phi}/2$. Using a 2-gluon wave
function of the form of eq.~(\ref{eq:wf}) yields the explicit form for the
binding potential  
\begin{equation}
V_{\Phi}(r)= \frac{\hbar^2}{2\mu_{\Phi}}\ \Big[\frac{\kappa}{a^2} (\frac{r}{a})
^{\kappa-2}\ [\kappa (\frac{r}{a})^{\kappa} - (\kappa +1)]\Big]+E_o\ .
\label{eq:5}
\end{equation}
To calculate the shape of the potential the used value of $E_o$ (which
gives only an absolute shift) is not important. With 
$\kappa$, $a$, $\delta m$ in table~1, slightly different solutions of
the binding potential were obtained, which
are given by the dot-dashed and dashed lines in fig.~4. Because of the
relation $2g\rightarrow (q\bar q)^n$ this potential can also be
considered as confinement potential between the emitted quarks. This
is in surprising agreement with the 1/r + linear form expected from
potential models~\cite{qq} and consistent with the lattice
data of Bali et al.~\cite{BBali}. It is important to note, 
that our potential~(\ref{eq:5}) reproduces the  1/r + linear
form without any assumption on its distance behavior; this is entirely
a consequence of the deduced radial form of the 2-gluon wave function,
see eq.~(\ref{eq:wf}). 

Absolute values of binding energies $E_i$ in the potential~(\ref{eq:4}) 
can be obtained by fitting the binding potential by a form 
$V_{fit}(r)=-\frac{a}{r}+b r$, and adding to the eigenvalues
$\epsilon_i$ a constant $\delta E$ to give a g.s.~energy
$E_o=\epsilon_o+\delta E$ consistent with $m_{\Phi}$. 
This yields bound state energies for the ground state and the
two lowest radial excitations given in table~1. 

By computing also eigenstates in the q-q potential~(\ref{eq:qq2}) we
find only one bound state with an energy in the order of -10 MeV. This very
low binding energy indicates clearly, that the glueball states must
have a large width, since $\Gamma\sim 1/E_o$ should be valid for the decay
width. This is consistent with the general expectation of
the width of glueball states ($\Gamma\ge$ 500 MeV,~see e.g.~ref.~\cite{MO}). 
From these results we may conclude, that the glueball ground state with 
$E_o$=0.68$\pm$0.10 GeV and a large width may be identified with the scalar
$\sigma$(550), which is most likely not a $q\bar q$ structure (see e.g.~the
review by Bugg~\cite{Bugg}).  
 
Glueball masses have been deduced also from lattice
simulations~\cite{glue,Bali}, in which a glueball mass below 
1 GeV has not been found. However, in these simulations $0^{++}$ glueball
masses have been extracted at about 1.7 and 2.6 GeV, which correspond
very nicely to the first and second radial excitation in table~1. Our
evidence for a low lying glueball is supported by QCD sum rule
estimates~\cite{Narison}, which also require the existence of a low lying gluonium
state below 1 GeV.
\vspace{0.2cm}

{\bf Conclusion}: We have presented a phenomenological 
method which yields a good understanding of important properties of the
non-perturbative structure of QCD. The basic assumption, that stable
2-gluon colour singlets exist, which decay into $q\bar q$ pairs, has allowed a
self-consistent determination of the densities of these 2-gluon
systems. These give important contributions to the gluon 2-point
functions, field correlators and gluon propagator, and yield a
two-gluon binding potential which can be identified with the
confinement potential determined from potential models and lattice QCD.  

A detailed comparison of our approach with lattice QCD will be important. In the
latter two-gluon fields are facilitated by loop integrations. By
relating gluon loops to our 2-gluon fields (recall that Wilson loops
are gauge invariant as colour singlet 2-gluon systems) one can
imagine, that Wilson's confinement picture (by evaluating closed gluon
loops around separated quarks) has a correspondence to our
interpretation of confinement as binding of 2-gluons. 

New information is obtained on the mass of glueballs. Consistent with
pole analyses of the gluon propagator and QCD sum rules we obtain a low mass of the
2-gluon system (glueball ground state) in the order of 0.7 GeV. This
is in the mass region of the scalar  $\sigma$(550). Higher radial
excitations are obtained at about 1.7 and 2.7 GeV, consistent with  $0^{++}$ 
glueball masses from lattice QCD. 
\vspace{0.2cm}

We thank P. Decowski, S. Krewald and N. Nikolaev for fruitful
discussions and M. Dillig for valuable comments and a careful reading
of the manuscript.  

\begin{figure}
\centering
\includegraphics [height=18cm,angle=0] {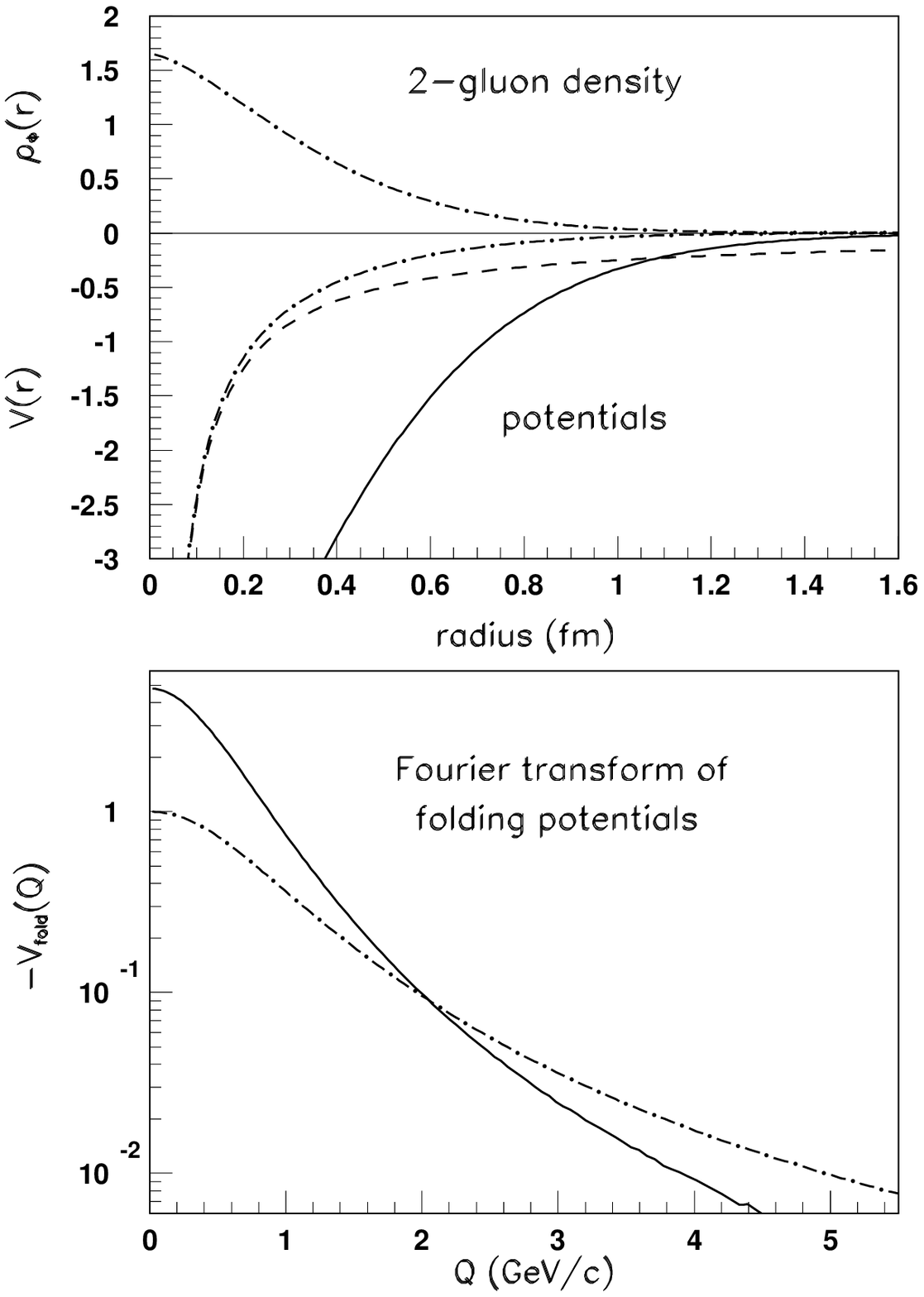}
\label{fig1}
\caption{Upper part: Two-gluon density $\rho_\Phi(r)$ (upper dot-dashed line)
  and potentials: 1-gluon exchange potential (dashed line), 
  modified by size effects (dot-dashed line) and folding 
  potential~(\ref{eq:qq2}) given by solid line. Lower part: Fourier
  transform of the folding potential using $\rho_\Phi(r)$ (solid line) and
  $\rho^p_\Phi(r)$ (dot-dashed line), respectively.}
\end{figure} 

\begin{figure} [ht]
\centering
\includegraphics [height=18cm,angle=0] {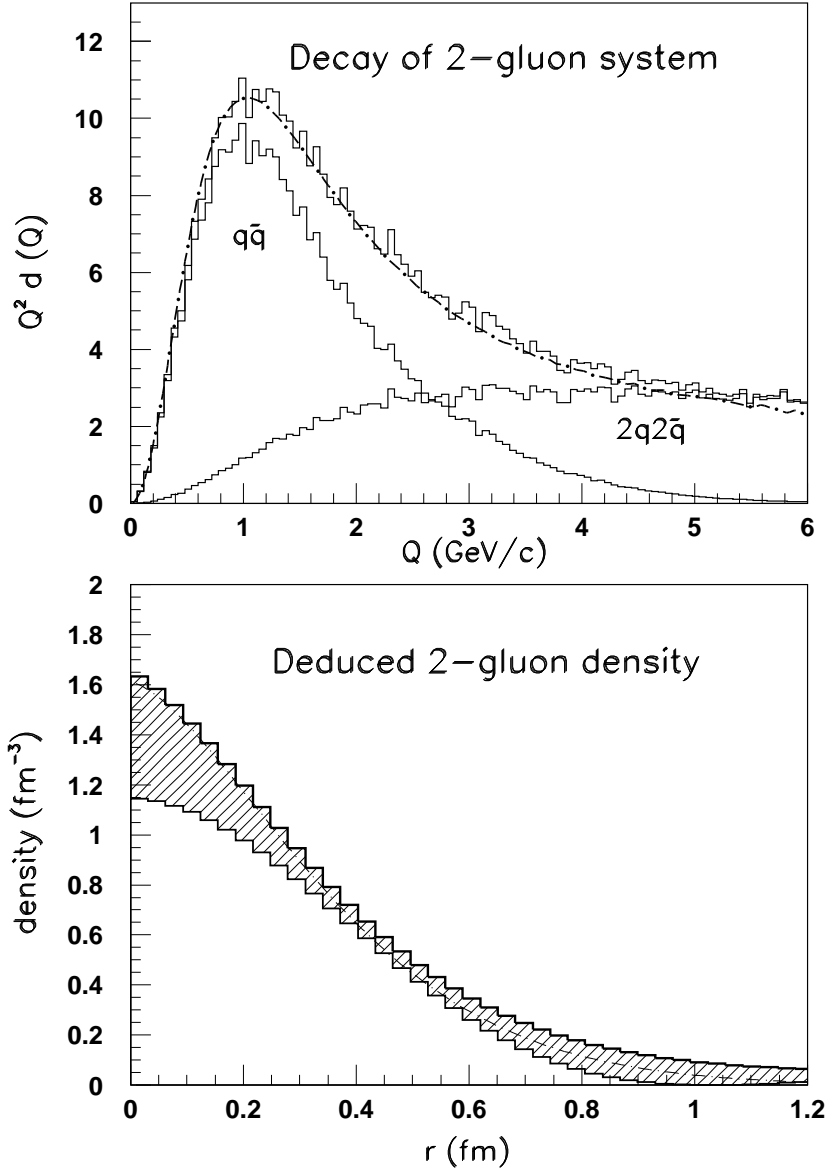}
\label{fig2}
\caption{Upper part: Resulting 2-gluon momentum distributions (multiplied by
  Q$^2$) for decay in $q\bar q$ and $2q2\bar q$ and sum.
Lower part: deduced 2-gluon density with estimated error band. The
dot-dashed lines represent solution~1 in table~1.} 
\end{figure}

\begin{figure} [ht]
\centering
\includegraphics [height=18cm,angle=0] {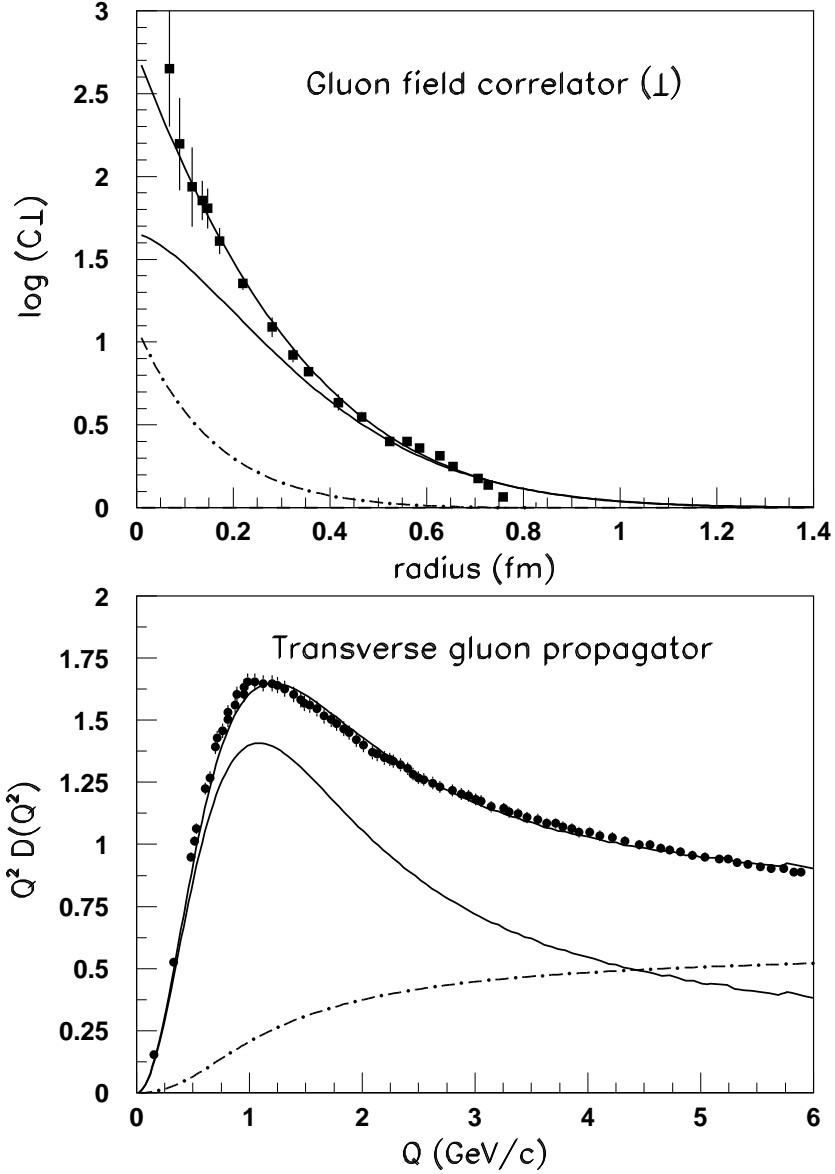}
\label{fig3}
\caption{Gluon field correlator $log(C_{\perp})$ from ref.~\cite{DiG} (upper
  part) and gluon propagator from ref.~\cite{Bow04} (lower part)
  from lattice QCD simulations in 
  comparison with our results. The lower solid lines correspond to the
  density $\rho_{\Phi}(r)$ and its Fourier transform $\rho_{\Phi}(Q)$,
  respectively, and the dot-dashed lines to an additional vector component
  with the same relative normalisation in the upper and lower part. The sum of
  both contributions yields a 
  good description of both lattice data.} 
\end{figure}

\begin{figure}
\centering
\includegraphics [height=14cm,angle=0] {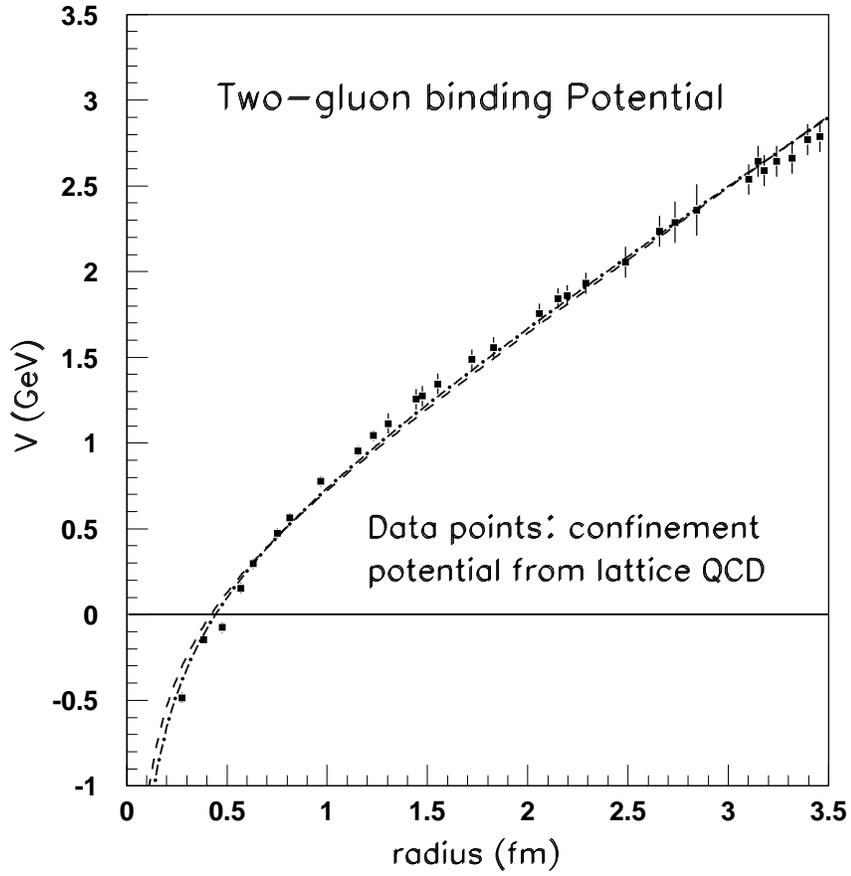}
\label{fig4}
\caption{Two-gluon binding potential~(\ref{eq:5}) given by the
  dot-dashed and dashed lines (with the respective parameters from table~1) 
in comparison with the confinement potential from lattice
calculations~\cite{BBali}. } 
\end{figure} 

\begin{thebibliography}{99}
\bibitem{GEN} A detailed write up of GENBOD can be found in the CERN library
\end{thebibliography}
\end{document}